# Comment to "Comment on 'Cartesian expressions for surface and regular solid spherical harmonics using binomial coefficients and its use in the evaluation of multicenter integrals'"

Telhat Özdoğan[1] and Metin Orbay[2]

[1]Department of Physics, Rize Faculty of Arts and Sciences, Karadeniz Technical University, Rize-Republic of TURKEY

[2]Department of Physics, Faculty of Education, Ondokuz Mayis University, 05189 Amasya- Republic of TURKEY

**Abstract**

The comments of Guseinov on our recent paper (*Czech. J. Phys., 52 (2002)1297*) have been analyzed critically. It is shown that his comments are irrelevant and also unjust. In contrast to his comment, it is proved that the presented formulae in our study are original and obtained independently, not by changing by the summation indices. It should be stressed that our algorithm is not affected from possible instability problems and also can be used in large scale calculations without loss of significant figures. Meanwhile, his comment on the transformation of our formula into his formula proves the correctness of our algorithm and therefore can be regarded as a nice sound of science.

The comments on our recent paper [1] by I.I. Guseinov have been analyzed critically. In his comment, he claims that the formulae we presented in Ref. [1] were not original and could be derived from his previous works by changing the summation indices. His claims will be replied items by items in the following:

***Comment 1*:** He claims that the formulae given by Eqs. (14-16, 21-23, 27-28) in our paper [1] are not original.

---

[1] Email: telhat@gmail.com
[2] Email: morbay@omu.edu.tr

*Reply to Comment 1:*

As is well known, spherical harmonics (SH) are the simplest and most familiar functions of mathematical physics, and also important for the calculation of molecular properties of atoms, molecules and solids, in Earth sciences, and in potential theory. The detailed discussion of the SH can be found in an excellent article by E.O. Steinborn and K. Rudenberg [2] (cited as Ref. [1] in our paper [1]).

As we understood from this comment, he either could not understand what the purpose of our paper is or behaved prejudicially as he always does.

It is seen clearly from the second paragraph of our paper [1] that the aim of our study was to give the cartesian expression for surface and regular solid spherical harmonics and to discuss the main advantage of our algorithm in the evaluation of the multicenter integrals. It is well known from the literature that the symmetry properties for spherical harmonics have not been introduced firstly by him as he claimed! and also was not the aim of our study. The symmetry properties presented for solid spherical harmonics was to test the expression presented in our paper [1].

**Comment 2**: He claims that the presented formulae for the generalized binomial coefficients $F_m(N, N')$ (Eq. (27) of Ref.[1]) and the expansion of the product of two normalized associated Legendre functions $T^{l1,l'1}(m,n)$ in ellipsoidal coordinates (Eq. (24 of Ref.[1]) are not original.

*Reply to Comment 2:*

As is well known in literature, the expansion coefficients of the product $(x+y)^m (x-y)^n$, called as generalized binomial coefficients $F_m(N, N')$, and the expansion of the product of two normalized associated Legendre functions in ellipsoidal coordinates can be found elsewhere (see: Eq. (A.12) of Ref. [3]; cited as. Ref. [15] in our paper [1]). The reference [3] is the fundamental book of quantum chemistry published many years ago before the work of Guseinov [4]. These expressions were applied in the calculation of multicenter molecular integrals over Slater type orbitals. So, these expressions are not introduced firstly by him as he claims !!!.

Contrary to his comments, we think that any physical or mathematical quantities obtained in two different ways can be transformed into each other. In this respect, the comments of Guseinov on the transformability of our formulae into his formulae prove the correctness of our algorithm and can be regarded as a sound and nice contribution to the science. Meanwhile, he published comments on some of our previous papers. Despite each of his comments has been replied, he ignores our replies and this causes misunderstanding for the readers and publishers. Replies to Guseinov's comments can be found from Refs. [5-9] or via the address http://www.geocities.com/telhat/telhatozdogan.html.